# Standardisation of practices in Open Source Hardware

Jérémy Bonvoisin[1], Jenny Molloy[2], Martin Haeuer[3], Tobias Wenzel[4,5+]

[1]University of Bath, Department of Mechanical Engineering, Bath, England;
[2]University of Cambridge, Department of Chemical Engineering and Biotechnology, Cambridge, England;
[3]Fraunhofer Institute for Production Systems and Design Technology (IPK), Berlin, Germany;
[4]Pontificia Universidad Católica de Chile, Institute for Biological and Medical Engineering (IIBM), Santiago, Chile;
[5]European Molecular Biology Laboratory (EMBL), Heidelberg, Germany
[+]Correspondence: jb2971@bath.ac.uk and tobias.wenzel@cantab.net

## Abstract

Standardising is an important component of the maturation of a field. It contributes to the formation of a recognisable identity and enables interactions with a wider community. This article reviews past and current standardisation initiatives in the field of Open Source Hardware (OSH). In open domains as OSH, standards are usually developed and implemented by the community itself. While early initiatives focused on aspects around licencing and intellectual property as well as documentation formats, recent efforts extend to ways for users to exercise their open permissions and to keep them findable and open in the digital infrastructure. We specifically introduce two standards that are currently being released and call for early users and contributors, the DIN SPEC 3105 and the Open Know How Manifest Specification. Finally, we reflect on challenges around standardisation in the community and relevant areas for future development such as an open tool chain, modularity and hardware specific interface standards.

## Introduction

Open Source Hardware[1] (OSH) is an emergent phenomenon applying to physical products an alternative approach to intellectual property (IP) that has proved to be relevant in Open Source Software (OSS) through decades of practice. As such, OSH faces the same doubts the OSS community did in its early days. Can the free revelation of intellectual assets lead to sustainable, functional products and economical development? In spite of its non-orthodox approach to IP, OSS has been the foundation of a highly innovative billion-euro economy and the basis of most digital products each of us use daily. The most iconic example of a successful open source software project is Linux. Since 2005, more than 13,594 people working for more than 1,340 companies have been involved in the development of the 22 millions lines of code constituting the Linux kernel. Since 2009, the development has been supported by a steady number of more than 200 companies (Corbet and Kroah-Hartman 2016). In 2019, Red Hat, one of the companies contributing the most to the development of the Linux kernel and flagging themselves as "provider of enterprise open source solutions", acquired a 3.4 billion $US revenue (Wonderlick and Walas 2019). These figures picture OSS as a concept that made their way from an alternative to a mainstream practice in a successful economic sector. They also underline that the complexity of OSS solutions can be comparable with those of proprietary software (approximated by the number of lines of code).

---

[1] in this article, we use the term Open Source Software/Hardware as a way to refer indifferently to open source and free software/hardware. This article does not intend to enter the debate about the differences between the *open source* and the *free* development philosophies.

For the younger phenomenon of OSH however, adoption by businesses remains low. The production of OSH has been mostly limited to non-commercial sectors such as grassroots communities, hobbyist markets, NGOs and academia (Troxler 2016). The complexity of hardware products developed under open source product development settings and released under open source licenses hasn't reached the level of those developed in proprietary settings. For example, in the automotive industry, it is common practice that development teams record over 20,000 CAD file changes a month (Audi, personal communication). A previous study by the authors investigated the number of file changes observable in the repositories of 105 OSH projects hosted on GitHub selected for representing the high end of OSH product complexity available. The maximum number of CAD file changes observed for each project over their complete lifetime was 7522, with a median value of 123 and average value of 509 (Bonvoisin et al. 2018). These figures indicate that "the level of maturity of Open Source Hardware (OSH) remains far lower than that of Open Source Software (OSS)" and raises the question whether "Open Source Hardware [will] follow the path of its sibling" (European Commission 2019).

When considering whether OSH will follow the same development path as OSS and whether an iconic project like Linux will ever emerge in OSH, the relative youth of the phenomenon has to be taken into account. With a 30-year delay compared to OSS, OSH is still a concept in the process of forming a consistent identity and a set of commonly-accepted best practices. This stands in contrast to the extensive and mature set of standards governing OSS. These are both *de facto,* i.e. adopted in practice by a community but not officially endorsed by a Standard-Setting Organisation (SSO), and *de jure,* which is defined by European standardisation legislation as "a technical specification, adopted by a recognised standardisation body, for repeated or continuous application, with which compliance is not compulsory" (European Union 2012).

The process of defining and standardising OSH is made difficult by the multifactorial and maybe ill-defined nature of *openness* (as discussed in Bonvoisin and Mies 2018 and; Balka, Raasch, and Herstatt 2014) and the consequent difficulty to draw a straight line between the practices referred to as OSH and other more or less related practices. A previous study from the authors published in this journal showed that hardware originators tend to interpret the concept of OSH differently when it comes to sharing hardware documentation (Bonvoisin et al. 2017). This spectrum of interpretations is further diluted by the "openwashing" discourses from industrial practice. Since openness became a trending concept, companies have been incentivised to identify with a set of principles they may not fully practice (L. F. Murillo 2017). While degrees of freedom are certainly necessary for a new phenomenon to emerge, agreement on practices and definitions is also necessary to give it momentum and maturity.

This article reviews the contribution of current standards to support OSH in its transit from a niche to a more mainstream position. It gives an overview of the benefits of existing standards and a preview of ongoing standardisation processes. In particular, we introduce two ongoing standardisation efforts, namely the DIN SPEC 3105 and the Open Know How Manifest Specification 1.0. We also identify gaps of existing standardisation efforts, calling for further initiatives.

## Definitions and dimensions of Open Source Hardware

Open Source Hardware (OSH), often simply referred to as Open Hardware, has only gradually evolved into a broad field now spanning machines and electronic boards, 3D printing as well as machine tools (predominantly desktop machine tools), vehicles (predominantly bicycles), robots, medical equipment, electricity production, agricultural machines, toys and games, optical products, musical instruments, clothing, and even materials, modified biological cell lines and so on (Bonvoisin et al. 2016).

The evolution of OSH definitions and standards can best be understood by having a look at what *hardware* means for OSS developers. Examples of OSH presented in the seminal book by Alicia Gibb "Building Open Source Hardware, DIY Manufacturing for Hackers and Makers" (Gibb 2014) are mainly electronic boards. In the early days of OSH, "hardware" was mainly understood as *electronic* hardware: the hardware on which OSS would run. In line with this, the first hardware–specific open source license -- the TAPR OHL -- was designed to protect digital telecommunication devices. Since then, the word *hardware* in OSH has acquired a broader meaning, encompassing diverse technologies beyond electronics, such as mechanics or textile. The TAPR OHL has become less popular as other licences emerged with more general utility such as the CERN OHL and SolderPad licence that cover a wider range of hardware types.

Mota's (2015) depiction of the early days of the open hardware movement mentions a few attempts to set standards that did not pass the test of time. The first of them was the Open Hardware Certification Program, announced in 1997, which understood openness as "availability of documentation for programming the device-driver interface of a specific hardware device" (Perens 1997). Over a decade later the Open Source Hardware and Design Alliance (OHANDA) formed in 2009 with the ambition to create a "certification model and a form of registration" (Neumann 2011). This initiative is remembered for its attempt to reformulate the four freedoms of the Free Software Definition (Free Software Foundation 2019) into the context of hardware. This reformulation was simplistic in the sense it was limited to changing the term "software" to "device" in the definition of the four freedoms. The reformulation did not state explicitly the possibility of making/replicating the device, which is an obvious limitation. The currently widely accepted definition and set of principles is provided by the Open Source Hardware Association (Open Source Hardware Association 2016), the first version of which arose following an Opening Hardware workshop in New York in March 2010 and has now reached version 1.0 (OSHWA 2020). The OSHWA Open Source Hardware Statement of Principles 1.0 covers the different dimensions of Open Hardware, which should be: transparent (right to study), accessible (right to modify and distribute), and replicable (right to make and sell).

The documents and definitions listed above cover some but, as we will further discuss in this article, not all of the dimensions that are often mentioned in the community as key to open hardware, such as modularity (to enable modification and reuse), interoperability (ability to combine different OSH parts), licencing, and openness through the entire product lifecycle (Oberloier and Pearce 2018; Booeshaghi et al. 2019; L. Murillo, Molloy, and Dosemagen 2017; Gavras 2018). These dimensions nevertheless guide various ongoing standardisation initiatives aiming to make open hardware projects recognisable and to aid their development.

## Licences, certification and documentation structure

Generally speaking, definitions enable the alignment of various standards and at a simplistic level provide a benchmark for compliance. However, they also have more complex roles to play in the development of future standards. As Powell (2015) points out, in terms of OSH licensing the OSHWA definition provides a important point of negotiation for OSH actors. In practice, OSH Definition-compliance is a key feature of any OSH licence (reviewed in Katz 2012) and for licences such as the CERN OHL that post-date the publication of the OSH Definition. The negotiation of the boundaries through the CERN OHL mailing list and other discussions was very active and covered both the importance of open hardware licenses to the OSH Definition in addition to the license structure and terms (Powell 2015).

Since 2018 OSHWA provides an OSH self-certification scheme based on transparent licensing of parts and products. As would be expected given the shared history and common originators of the initiatives, the OSHWA certification is built around the OSH Definition and "makes it easy for creators and users to identify hardware that meets the community definition of open source hardware". As a self-certification process that is broadly focused on licensing and aspects of intellectual property, this initiative does not contribute to standardisation of other dimensions of OSH mentioned in the previous section. For example, OSHWA members may review the quality of documentation of submitted projects but the criteria used are not formalised, leaving a gap for further standards and certification development.

A particularly active area to date has been into the standardisatoin of documentation formats. For example, oManual is an XML standard for documentation and manuals that was approved as the IEEE 1874 Standard for Documentation Schema for Repair and Assembly of Electronic Devices in 2013 (oManual 2020; IEEE 2014), and subsequently integrated into a commercial documentation platform[2]. It is also compatible with the more general Darwin Information Typing Architecture standard. oManual had the goals of enabling portability through an XML schema as well as readability through the integration of annotated images and a linear narrative. It was not designed specifically for OSH and its linearity and the fact that all current implementations are proprietary have limited its use in the OSH community. DocuBricks[3] is a documentation standard devised specifically for OSH and with a focus on modularity and interoperability - thus enabling different "bricks" with files and instructions to be combined into one project even with conflicting share-alike licenses. One goal is to encourage functional explanation of modules (bricks) by explicitly naming them in a project. It is also based on an XML schema, but is incompatible with the IEEE 1874 Standard due to the divergence in design goals. Git Building[4] is a recent initiative, yet without major release, which aims to integrate Open Hardware documentations with Git-based version control repositories. Relying on Markup language, it is easily human readable without additional software, at the expense of modularity and machine readability. Due to Git Building's resemblance of the DocuBricks layout and Bill of Materials (BOM) management, the two formats may be convertible when structured accordingly. Major documentation hosting platforms used by OSH projects do not currently use a documentation or content standard, which may either be due to a lack of development effort or due to the preferences of the largely hobbyist-focused community.

## Transparency through enforceable content - the DIN SPEC 3105

So far, we can see that efforts to shape the contours of what "open source hardware" means have primarily focused on legal aspects of OSH or documentation schema. They haven't succeeded in establishing a clear OSH identity based on sharp definitions and enforceable compliance i.e. precise and objective criteria delineating what OSH is from what it is not (Bonvoisin et al. 2017). Such criteria are required for OSH practitioners to establish a consistent public discourse about OSH and a general sense of trust required by businesses to take up the concept and scale it up to a wider level. This is the issue DIN SPEC 3105 "Open Source Hardware" addresses.

DIN SPEC 3105 "Open Source Hardware" is a standardisation effort initiated by Open Source Ecology Germany e.V. and officially launched by the German Standardisation Organisation DIN e.V. (Deutsches Institut für Normung e.V.) April 2019 (DIN 2019). The project was led under DIN's "Publicly Available Specification" procedure (PAS) delivering a public document that can "be used as a basis for a full standard". It involves "smaller, more agile working groups" and is an "effective marketing instrument [...] widely accepted by customers and potential partners alike". It is the first *de jure* rather than *de facto* standard developed for OSH.

---

[2] https://www.dozuki.com/, accessed 12/04/2020.
[3] https://www.docubricks.com, accessed 12/04/2020.
[4] https://gitbuilding.io, accessed 12/04/2020.

This standard contains two parts:
- DIN SPEC 3105-1 "Open Source Hardware - Requirements for documentation"[5] aimed at delivering an unambiguous definition of the term Open Source Hardware based on objective and enforceable criteria.
- DIN SPEC 3105-2 "Open Source Hardware - Community-based assessment"[6] builds upon the definitions provided by DIN SPEC 3105-1 to define requirements for an assessment procedure for OSH products based on reviews by OSH community members -- emulating the model of peer-review used in scientific publishing.

**DIN SPEC 3105-1 - Requirements for documentation**

The starting point of the DIN SPEC 3105-1 is that the applying reference definition of the term "Open Source Hardware" (OSHWA's definition mentioned earlier in this text) only focuses on the licensing aspects of product-related information disclosure and does not set concrete requirements regarding the content of the information to be disclosed. DIN SPEC 3105-1 aims at "extending the OSHWA Definition 1.0, which itself extends the Open Source Definition".

This document makes significant original contributions to clarify the concept of OSH and addresses several pitfalls of previous standards:
- It breaks down the requirement made by the OSHWA Definition 1.0 to enable "anyone [to] study, modify, distribute, make, and sell the design or hardware" into clear, actionable ways to meet that requirement. Therewith, it defines four "Rights of Open Source" transposing the "four essential freedoms" stated in the Free Software Definition (Free Software Foundation 2019) and links them with concrete requirements in terms of documentation content. These rights are the right to study, to modify, to make, and to distribute.
- It acknowledges that OSH is not only a matter of licensing but also a matter of documentation contents. It differentiates between *granting* the four rights of open source and enabling others to effectively *exercise* these rights[7].
- It acknowledges that the content of the documentation to be shared depends on the hardware technologies embedded in the product under consideration (e.g. electronic hardware). To account for these dependencies, the specification only dictates generic requirements in terms of documentation contents. These contents are in turn specified by so-called "Technology-specific Documentation Criteria" which are specific for each technology.
- It acknowledges that the content of the documentation also depends on the audience targeted by the documentation. It is not practicable to allow "*anyone* [to] study, modify, distribute, make, and sell the design or hardware" since not everyone has the same capacity to read and make use of product documentation. Instead, documentation is required to address a defined group of *recipients*. The default is that documentation must provide no less information than what specialists in the fields of this technology would require to exercise the four rights of open source hardware.
- It adopts a product life cycle perspective, considering that "making" a product is not only about manufacturing it, it is also about using, maintaining, updating and processing it at end-of-life. It defines documentation as information allowing to operate all activities belonging to the product life cycle, from raw material extraction to end-of-life. Therewith, it drops the arbitrary and artificial barrier made by previous standards between production

---

[5] https://gitlab.com/OSEGermany/OHS/-/blob/master/DIN_SPEC_3105-1.md
[6] https://gitlab.com/OSEGermany/OHS/-/blob/master/DIN_SPEC_3105-2.md
[7] to give simple illustrations of the difference between granting a right and enabling people to exercise this right: you may have the right to fly to the moon but may not have the effective possibility to do so; you may be entitled by law to tax benefits but may not have the capacity to overcome the administrative hurdles to claim these benefits.

and the rest of the life cycle.

### DIN SPEC 3105-2 - community-based assessment

DIN SPEC 3501-2 intends to frame the application of the DIN SPEC 3105-1 in practice by defining a dedicated assessment procedure to be implemented by any willing conformity assessment body, who will play a role akin to that of a scientific journal editor: to moderate a constructive discussion between authors and reviewers and take a publication decision based on the convergence of this discussion. The underlying assumption is that, for the specific context of OSH, peer-certification is a more trustable assessment system than self-certification and a more practicable assessment system than third-party certification. While self-certification does not provide enough transparency to check compliance, third party certification implies dependences to certification bodies and costly certification processes. By vesting responsibility and decision making power in the community who understand the social and technical norms of OSH, the trust that OSH developers hold in the credibility and validity of the standard may be higher.

### Open access and collaborative future development

The first release of DIN SPEC 3015 was published in April 2020. It specifies requirements for technical documentation and compliance requirements beyond currently existing definitions and standards. Therewith, it clarifies what OSH means and contributes to identity building in the OSH community. This unavoidably goes hand in hand with higher barriers to adoption. The future will tell whether this standard found the right balance between rigour and flexibility and to what extent OSH development communities are ready to invest in documentation efforts to comply with the ideals of open source.

An aspect potentially playing a role in adoption is the public availability of the standard: The DIN SPEC 3105 is the first German standard to be published under an open license (CC-BY-SA 4.0) and to implement an open standardisation process. Viewing and using the standard is thus possible without costly licenses or permission from DIN e.V. or any further restrictions apart from copyleft and DIN's trademark protection. The open process also provides a mechanism for the OSH community to adjust the standard to their needs in future versions. The standard (and its possible derivatives) can be field-tested by users and then adjusted by collectively submitting a new version to DIN e.V. in order to update the official DIN SPEC 3105 release. This process aims to 1) attract and resolve much more feedback compared to the conventional standards process, 2) point out gaps where new standards are needed and 3) provide a fast and flexible way to test and adjust standards before an official release.

## Discoverability through Metadata - the Open Know-How Manifest Specification

While DIN SPEC 3105 addresses the content of OSH documentation, including a minimal set of metadata for a documentation release, it remains platform-agnostic and does not adopt any formal metadata and data schema. The question of metadata schema is addressed by another initiative bringing an additional layer of standardisation to facilitate sharing open hardware documentation: the Open Know-How Manifest Specification.

The Open Know-How Manifest Specification is an initiative from the Open Know-How Working Group, who focuses on standardised protocols for the exchange of Open Hardware related information online. The initiative is based on the observation that Open Hardware requires new networks of tools and information in order to ensure that users can find and access appropriate hardware designs, local physical tools to make them, and potentially customers to sell units to. It was established in 2019, bringing together a number of members from academia, NGOs, companies and OSH community groups with the objective of issuing standards to address three levels of sharing documentation information (Know-How) on OSH:
1. discoverability, i.e. the ability to find the documentation regardless of where it resides on the internet, in part through making documentation accessible to web crawlers;
2. portability, i.e. the ability to share documentation on different platforms or to transfer it from one platform to another;
3. platform interoperability, i.e. the ability to update, relate and join documentations across different online repositories and documentation formats.

The first standard released by the working group is the Open Know-How Manifest Specification 1.0[8] addressing discoverability (Open Know How Working Group 2019). The working principle of this standard is creating an additional file, called a "manifest", for each Open Hardware documentation, which contains metadata relative to the documentation and to other files commonly shared next to hardware documentation, e.g. the applying license terms and the "readme" file. The standard intends to promote discoverability of open source hardware by establishing a machine-readable format which enables search engines and web crawlers to index open source hardware documentation metadata from the internet. A demonstrator of such an index is provided by the working group. It gathers metadata from approx. 400 pieces of hardware and make them available in a single URL[9]. The manifest file may be written by hand by the authors or it may be auto generated by documentation repositories if the corresponding documentation is sufficiently structured, e.g. in the DocuBricks format, to assess its content automatically. The manifest file links to the documentation and can be thus placed anywhere online, in the documentation repository or in another dedicated location.

The standard delivers a template formatted in YAML markup language (Ben-Kiki, Evans, and döt Net 2019) which is both machine-readable and easy to modify for non-expert users. The manifest file covers the following metadata:
- basic information such as contact person, licence, language and the locations of (i.e. links to) documentation, bill of materials and project homepage
- descriptive information such as intended use, keywords and development stage
- locations of (links to) some more advanced information such as risk assessments, quality control, or maintenance protocols;
- the ability to make basic relationships between OSH items more transparent e.g. by using "version", "variant-of" and the relationship term "derivative-of" to indicate that the design is derived from another and providing a link to that documentation.

This content makes manifest files fully compatible with the requirements from DIN SPEC 3105. The project has set a goal of 500 OSH projects adopting the Open Know-How Manifests by end of 2020 and is actively recruiting projects to publish a manifest and index it in their demonstrator[10]. Like DIN SPEC 3105, the standard has been a collaborative effort. It has open channels on the open standard editing platform StandardsRepo[11] where feedback and improvements can be suggested by raising an issue or proposal.

The manifest may be the first a series of standards from the Open Know-How Group aimed at building an "Internet of Production", a network enabled by open exchange protocols to locally share and access Open Hardware documentation, as well as needed tools and the information how to use those. This essential infrastructure for Open Hardware may be aided by the creation of appropriate standards which ensure that diverse projects and platforms can exchange information effectively, and to avoid potential vendor lock-ins for the online infrastructure on which the community relies.

---

[8] https://app.standardsrepo.com/MakerNetAlliance/OpenKnowHow/wiki, accessed 12/04/2020.
[9] https://search.openknowhow.org/, accessed 12/04/2020.
[10] https://github.com/OpenKnowHow/okh-search/blob/master/projects_okhs.csv, accessed 12/04/2020.
[11] now called Barbal https://barbal.co, accessed 12/04/2020.

## Discussion and outlook

Together with existing standards providing appropriate licensing terms, the ongoing attempts to define open hardware documentation content and discoverable, interoperable metadata provide a basis for building transparent certification processes and trust around the concept of OSH for a wider audience. Nonetheless, they still only cover some of the aspects of the multifaceted concept of hardware openness as encountered within OSH communities of practice. As Gavras (2018) highlighted: both licensing terms and documentation contents are not "structural properties of the source code"; they are not "inherent properties" of a given piece of hardware that is a candidate for the OSH label. Neither completeness of documentation nor the licensing terms under which the documentation is released say anything about the piece of hardware itself. This raises questions on whether there are inherent properties of hardware that make it more or less amenable to standardised practices and to what extent there are gaps that are not addressed by existing efforts.

### Modularity

Modularity is one of the inherent properties of hardware which is often mentioned as a key enabler in hardware openness. The appetite for modularity in OSH communities is reflected in the lego-like design style of some iconic projects like Openstructures, Wikihouse or XYZ CargoBikes, and which is more generally characteristic of the DIY movement which shares overlapping values with the open source movement. One of the arguments behind the importance of modularity in open source is that it is "a form of task decomposition" (Dafermos and Söderberg 2009) and therewith influences organisation of production and development. For Kostakis and Papachristou (2014), modular product design is one of the key conditions for the emergence of commons-based peer production. Product structure and organisational structure of the development team also go hand in hand, as stated in the often-cited "mirroring hypothesis", according to which "organizations which design systems are constrained to produce designs which are copies of the communication structures of these organizations" (Conway 1968). McCormack (2012) empirically confirmed that modular design is characteristic of the FOSS development model in general and that products developed by loosely-coupled organizations are more modular than those developed by tightly-coupled organizations. It is reasonable to expect such correspondence between product and organisational structure in OSH. The influence of product structure on the product development process is of importance considering that the very concept of open source is generally interpreted as a product development model, and not only as an intellectual property management scheme (see for example Gacek and Arief 2004; Raasch and Herstatt 2011; Moritz, Redlich, and Wulfsberg 2018).

However important the concept of modularity may be, it may not be sensible to integrate it as an enforceable aspect of the OSH definition, partly because there is no commonly accepted definition of what it means for a product to be "modular" and of how to quantify product modularity (see a discussion of this question in Bonvoisin et al. 2016). Nonetheless, there is certainly value in collating further guidance for practitioners on how to leverage modular product design for reaching means of distributed development and production as well as product maintainability and upgradability, among others. Enabling a modular documentation of hardware may also support practitioners in designing modular products. Some open hardware sharing platforms already allow to define modules that build up into a full documentation through a hierarchical structure, e.g. Wikifactory, and XML specifications like DocuBricks were predicated on the need for modules and submodules that could be assembled into different configurations.

### Interoperability

Another property of hardware that is not reflected in the standards discussed above is interoperability, or the ability for a product to interface with others in an ecosystem. Interoperability is necessary to enable modular design and to realise some of the most commonly cited advantages of OSH. There, however, are complications including interoperability within general OSH standards. For example, while the standards described in this article cover in principle all open hardware projects, interoperability relies on a network of technology-specific standards for communication, software and data exchange between different components and devices, in addition to physical interoperability of components such as dimensions and thread size.

One example of such technology specific standard is the BioBrick (Shetty, Endy, and Knight 2008). It is a standard in the area of synthetic and molecular biology for the physical composition of genetic parts. The BioBrick can be counted towards technology specific Open Hardware standards since it is concerned with physical material (DNA), even if not of classical electro-mechanical kind. It specifies BioBrick "parts" as functional genetic sequences (genes) that use a defined set of sequences to interface with BioBrick "vectors" as the structural DNA backbone used to handle genes for cloning. The interface sequences that flank the genes are recognised by a defined set of restriction enzymes, which are needed to assemble and transfer parts in a universal manner.

Instead of generalistic standards, general recommendations could be made to adopt domain-specific open technical standards in OSH designs where possible but the term "open standards" also has multiple meanings (Larrouche 2014). Some, like the DIN SPEC 3105, i) arise from an open community process; ii) may be accessed and read by anybody and iii) may be used by anybody without royalty payments, which is aligned with OSH values and practices. Others are open solely in one of these areas and currently the nature of hardware is such that most OSH designs are not using open technical specifications at every level of their design. In contrast, many make use of common but proprietary electronics architectures and interfaces e.g. Bluetooth. A major area of growth and funding in open hardware are open Instruction Set Architectures (ISAs) such as RISC-V, where the RISC-V Foundation manages the standard specifications and RISC-V software is managed by respective open source software projects; these initiatives allow users to design interoperable microprocessors that can be hard-wired or programmed (for example in field programmable gate arrays - FPGAs). Open ISAs increase opportunities for OSH projects to adopt open approaches at multiple levels of design but the existing specifications do not address the desirability or enforceability of such decisions.

### OSH Development Tools

Beyond the interoperability of the hardware itself, there are decade-long issues of interoperability in product development designs and data, which the initiatives discussed in this paper do not solve, despite setting new standards in the sharing of hardware documentation. The poor interoperability of Computer-Aided Design (CAD) software and the closedness of their output file formats is an aspect that plagues data sharing in proprietary product development and also affects OSH. Standardisation efforts such as the CAD file format STEP (ISO 1994) brought practical but limited solutions to this problem. It is unlikely that the OSH movement will play a driving role in solving this complex issue among the incumbent proprietary software providers, as they have established markets and benefit from a degree of vendor lock-in with major corporate customers. However, a workaround to this deadlock is the development of FOSS for hardware development, such as mechanical or electronic CAD software and applications such as FreeCAD, OpenSCAD and KiCAD are gaining users and functionality in these respective fields. While there is still a long way towards a fully open source toolchain competing with the capabilities of proprietary software, great progress has been made in the last years and this may need to be considered in future standards specifications.

### Sharing materials and tools

Another important aspect of Open Hardware is accessing material and tools required to build or reuse respective products. This is a problem unique to hardware, as the hardware itself cannot simply be downloaded for use on largely interoperable personal computers or single machines able to handle all aspects of digitally manufacturing most hardware. How these materials and tools can be more effectively shared is an important question also for the participation of resource poorer areas in the world, where local solutions might have an even larger impact, but e.g. the density of community workshops such as Maker Spaces and FabLabs are lower, as these are currently mainly concentrated in the US and Europe[12]. The recent Open Material Transfer Agreement (Kahl et al. 2018) is the first material transfer agreement that fulfills minimal OSH community requirements. A related initiative to Open Know-How, is the Open Know-Where initiative still in its formation, which aims to help the community in locating tools or manufacturers for open tools close to their location in a distributed manner.

### Hosting standards

Another future question is where to host community standards. Currently, most community-born standardisation initiatives are hosted by a community organisation website or the website of the development community without formal affiliation. In the future, it might be that standards organisations such as DIN open up further towards hosting community discussion and standards releases, and also the IEEE standards association considers making its platform more amenable to less formally organised communities for this purpose (IEEE 2020). The experience gained in establishing the DIN SPEC 3105 will also inform DIN e.V. 's decision on further opening of their standardisation processes. It provides a practical example to other standards bodies of how they might effectively collaborate with open source communities using open approaches.

## Conclusion

We have outlined how recent standardisation initiatives for practises in Open Source Hardware have widened the scope beyond the previous focus topics of definitions, licences, and documentation formats. In particular, the new standard DIN SPEC 3105 provides enforceable criteria for documentations, the "source code" of hardware, validated in a community-based assessment process. With this approach, it aims to improve the recognisability of OSH and build a bridge connecting industry, science and the worldwide open source hardware community. The Open Know-How Manifest Specification 1.0, the other recent standard we highlight in this review, specifies a file or "manifest" to be created in addition to documentation, which is machine readable and contains project metadata and references to the documentation. This file makes OSH documentations discoverable online independently of their host repository. Importantly, both initiatives have brought in various stakeholders in the Open Hardware community to contribute to their development and ensure early acceptance and adaptation. Both initiatives are open for anyone's active contribution for future versions.

Two further important dimensions of Open Hardware are modularity and interoperability, including software and data aspects. These dimensions are, however, more difficult to generalise and to evaluate in terms of enforceable criteria. They are more amenable to standardisation in their respective OSH sub-communities at the sector/domain level. Hopefully for future standards, we will see increased activity from such communities, and the spread of this type of discussion from the global community to subject groups. This process may involve subject journals which are already increasingly publishing Open Hardware manuscripts and Learned Societies could play an active role in standardisation efforts.

---

[12] as observable from the data from https://www.atlasofinnovation.com, accessed 12/04/2020.

An open infrastructure for OSH designs is another key enabler for users to effectively exercise the rights granted under OSH licences. However, in particular the Open Source toolchain for hardware design is beyond the control of the OSH community and unlikely to become more interoperable soon. A good way forward seems to be the continued support of Open Source CAD software. A consistent and powerful open toolchain would likely also change the criteria of current OSH standards towards more stringent policies. Open infrastructure may furthermore extend to cover network-access to physical tools and platforms for community standards development itself.

Finally, we shall be reminded that all the covered initiatives are essentially voluntary open source communities, trying to make Open Hardware development easier and better, while averting Open Washing or commercial lock-ins. These initiatives rely on our support by using and referencing them. They also rely on feedback in order to find the right balance between efficient requirements and creative deregulation in a fast changing field. While we are working on further standards, we should remind ourselves that the community grows only if we build on each other's work, and that this development is only as inclusive as underrepresented communities are taken on board, so that newly developed standards work for all. We certainly need further growth in participation, efficiency, and adaptation to fulfill our vision of ubiquitous Open Hardware products and large open project collaboration in the image of our sibling Open Source Software, which enjoys a 30-year head start. Active usage and participation wanted!

## Acknowledgments

This article is published with the support of the European Union's Horizon 2020 research and innovation programme under grant agreement no. 869984.